Inferring, comparing and exploring ecological networks from time-series data through R packages `constructnet`, `disgraph` and `dynet`

Anshuman Swain, Travis Byrum, Zhaoyi Zhuang, Luke Perry, Michael Lin and William F Fagan

Department of Biology, University of Maryland, College Park, MD 20742, USA

Summary:

1. Network inference is a major field of interest for the ecological community, especially in light of the high cost and difficulty of manual observation, and easy availability of remote, long term monitoring data. In addition, comparing across similar network structures, especially with spatial, environmental, or temporal variability and, simulating processes on networks to create toy models and hypotheses – are topics of considerable interest to the researchers.

2. A large number of methods are being developed in the network science community to achieve these objectives but either don't have their code available or an implementation in R, the language preferred by ecologists and other biologists.

3. We provide a suite of three packages which will provide a central suite of standardized network inference methods from time-series data (`constructnet`), distance metrics (`disgraph`) and (process) simulation models (`dynet`) to the growing R network analysis environment and would help ecologists and biologists to perform and compare methods under one roof.

4. These packages are implemented in a coherent, consistent framework – making comparisons across methods and metrics easier. We hope that these tools in R will help increase the accessibility of network tools to ecologists and other biologists, who the language for most of their analysis.

**Introduction**

The usage of networks in ecology and allied fields has increased considerably in the past two decades, owing to the explosion of new tools and theoretical findings in network science (Poisot et al., 2016). The entities represented by ecological networks can be quite varied – they can be animal social interaction networks (Croft et al., 2008; Krause et al., 2015), species interaction networks (trophic networks, mutualistic networks, parasitic networks, antagonistic networks) (Dunne et al., 2002; Pascual and Dunne, 2006; Kefi et al., 2012; Bascompte and Jordano, 2013; Shaw et al., 2021), or ecosystem flux models (Proulx et al., 2005). The study and exploration of these networks have influenced our understanding of communities and ecosystems in a tremendous way, allowing for a holistic representation of ecological systems (see reviews: Delmas et al., 2019; Guimarães, 2020).

The underlying data from which a network can be constructed or inferred can come from a variety of sources. However, the usefulness of a network is only as good as the data from which it is constructed (Poisot et al., 2016). Direct field observations and lab experiments form a large portion of the data underlying ecological networks, but such data are usually either difficult to obtain, expensive to record, or require large amounts of manual effort (see Jordano, 2016; Poisot et al., 2016). An alternative way to construct networks is to infer interactions and associations from existing records of direct interactions or indirectly through numerous ecological and biological traits (see Bartomeus et al., 2016). One specific type of data that has become relatively important in network reconstruction is time-series data (Runge et al., 2019), especially due to the availability of inexpensive continuous monitoring methods and/or the ease of access through digitization of long-term ecological records from various locations around the globe (see Lindenmayer et al. 2012; Shin et al., 2020). Temporal data, through the lens of network science, have particular advantages as they are unidirectional and provide a basis for detecting causal mechanisms in ecological dynamics (Sugihara et al., 2012; Dornelas et al., 2013; Adams et al., 2020), especially where manipulation experiments are not feasible (Bálint et al., 2018), and in paleo-ecological analysis, where the ecosystems in question no longer exist (Muscente et al., 2018; Swain et al., 2020).

Although numerous works have been dedicated to network inference from time-series data, selecting and comparing among appropriate techniques has always remained a prime problem (Runge et al., 2019). Chiefly this is because: (1) methods are based on different assumptions about the system, and a reference list is often not found in a single location for comparison (McCabe et al., 2020), and (2) methods are implemented in different ways and not all of them are usually found in the same programming language and implementation format (especially in R – the language preferred by most ecologists).

Here we address both of these problems by bringing together numerous network inference methods (in package *constructnet*) and network distance metrics (in package *disgraph*), in a unified implementation pipeline in the R statistical language. The distance metrics can be used to compare networks constructed from different methods using the same time-series, and also different (but similar) networks constructed from different time-series but using the same method. In addition, we also provide numerous methods to simulate time-series data using specific simulation processes on already known/inferred networks (through the package *dynet*). Brief summaries of the methods, distances and simulation models are provided in the sections below.

`constructnet`: **Network inference methods**

For inference methods, the input will be a time-series matrix (parameter `TS`) with variables of interest as columns and time-stamped observations as rows. The main output will be a graph object, and an adjacency matrix.

| Method (and function name) | Brief description, with additional parameters (APs) | Selected references |
|---|---|---|
| Convergent Cross Mapping (`convergent_cross_mapping_fit`) | Based on dynamical systems theory, this tool can be applied to systems where causal variables have synergistic effects in order to detect causality (**APs:**`- tau: Number of time steps for a single time-lag`) | Sugihara et al., 2012 |
| Correlation matrix (`correlation_matrix_fit`) | Pairwise correlation with regularization and thresholding (**APs:**`- num_eigs: The number of eigenvalues to use`) | Friedman and Alm, 2012 |
| `Free Energy Minimization` (`free_energy_minimization_fit`) | Decouples iterative model updates from goodness of fit to enable the usage of goodness of fit as a natural rationale for estimating optimality, thereby avoiding overfitting. | Hoang et al., 2019 |
| Granger Causality (`granger_causality_fit`) | Estimates interactions through pairwise and directional temporal predictability (**APs:**`- lag: Time lag to consider`) | Guo et al., 2008 |
| Graphical LASSO (`graphical_lasso_fit`) | Attempts maximum likelihood estimation, as per a Gaussian base model, and regularization by imposing a penalty on the strength of 'interaction' and number of 'interactions' (**APs:**`- alpha: coefficient of penalization, higher values mean more sparseness; max_iter: maximum number of iterations; tol: stop the` | Friedman et al., 2008 |

| | algorithm when the duality gap is below a certain threshold) | |
|---|---|---|
| Marchenko-Pastur (marchenko_pastur_fit) | Performs 'noise removal' from a correlation matrix to create a signed network (**APs:**- remove_largest: If True, the eigenvector associated to the largest eigenvalue is going to be excluded from the reconstruction step; metric_distance: If True, the correlation is transformed by defining a metric distance between each pair of nodes; tol: tolerance parameter) | Bonanno et al., 2004 |
| Maximum Likelihood Estimation (maximum_likelihood_estimation_fit) | Uses maximum likelihood estimation to infer interaction strengths (**APs:**- rate: rate term in maximum likelihood; stop_criterion: if True, prevent overly long runtimes) | Snijders et al., 2010 |
| Mutual Information (mutual_information_matrix_fit) | Relies on applications of information theory for quantification of the 'amount of information' obtained about one variable through observing others (**APs:**- nbins: number of bins for the pre-processing step) | Bianco-Martinez et al., 2016 |
| Thouless-Anderson-Palmer (thouless_anderson_palmer_fit) | Uses the Thouless-Anderson-Palmer (mean field) approximation to simplify interaction assumptions among variables. Performs well for large sample sizes and small variability in interaction strengths | Hoang et al., 2019 |

`disgraph`: **Network distance metrics**

For distance metrics, the input will be two graphs (igraph objects) or their associated matrices. The output will be their graph distance.

| Distance metric (and function name) | Brief description, with additional parameters (APs) | Selected references |
|---|---|---|
| Frobenius (`dist_frobenius`) | Distance measured between two (network) adjacency matrices a and b as $\sqrt{\sum_{i,j}\left|a_{ij}-b_{ij}\right|^2}$, where i and j are row and column indices | Deza and Deza, 2009 |
| Hamming-Ipsen-Mikhailov (`dist_hamming_ipsen_mikhailov`) | Combines the more commonly used Hamming distance and the Ipsen-Mikailov distance (see below) through a Cartesian product to create a (metric) space for calculating a Euclidean metric (**APs:**- `combination_factor`: Numeric factor to be combined with IM metric; `results_list`: Logical indicating whether or not to return results list) | Jurman et al., 2015 |
| Ipsen-Mikhailov (`dist_ipsen_mikhailov`) | Compares the spectra of two network-associated Laplacian matrices (**APs:**- `hwhm`: Numeric parameter for the Lorentzian kernel; `results_list`: Logical indicating whether or not to return results list) | Ispen, 2004 |

| Laplacian Spectral (dist_laplacian_spectral) | Quantifies the topological distance between two networks based on the contrasting spectra of their (normalized) Laplacian matrices via both Euclidean and Shannon-Jensen measures (**APs:**` - normed: Logical parameter for the Lorentzian kernel; kernel: Character indicating kernel type. Can be "normal", "lorentzian", or NULL; hwhm: Numeric indicating half width at half maximum; measure: Character indicating measure type. Can be "jensen-shannon" or "euclidean";k: Numeric indicating number of eigenvalues kept, NULL means all; which: Character prioritizing which eigenvalues are kept, see get_eigvs below for more; results_list: Boolean indicating whether to return results_list`) | Banerjee, 2012 |
|---|---|---|
| Polynomial Dissimilarity (dist_polynomial_dissimilarity) | Takes polynomials linked to the eigenvalues of (network) adjacency matrices into account for distance computation (**APs:**`- k: Numeric maximum degree of the polynomial; alpha: Numeric weighting factor`) | Donnat and Holmes, 2012 |

`dynet`: **Process simulation on networks**

For process models, the input will be an adjacency matrix of a graph (square matrix; parameter: `input_matrix`), length of the desired time series (parameter: `L`). The output will be a generated time-series.

| Model (and function name) | Brief description, with additional parameters (APs) | Selected references |
|---|---|---|
| Ising-Glauber process (`simulate_ising`) | Simulates the Glauber process in the Ising model, where each network node has two possible states, and switches between them based on the degree of the node, number of active neighbors, and a tuning parameter<br>(**APs:**- `init: vector initial condition, which must have binary value (0 or 1) and must have length N; beta: Inverse temperature tuning the likelihood that a node switches its state. Default to 2.`) | Castellano and Pastor-Satorras, 2006; Gleeson et al., 2013 |
| Kuramoto process (`simulate_kuramoto`) | Kuramoto oscillator system simulates synchronization processes, where each node in a network, which possesses an internal frequency, synchronizes its frequency with others based on neighbor interactions and a tuning parameter<br>(**APs:**- `dt: size of timestep for numerical integration; strength: coupling strength (prefactor for interaction terms; phases: vector of of initial phases; freqs: vector of internal frequencies`) | Rodrigues et al., 2016 |

| Lotka-Volterra dynamics (`simulate_lotka`) | Simulates the Lotka-Volterra competition model with interspecific and intraspecific interaction strengths modeled by the network edge weights (**APs:**`- init: Initial condition vector; gr: Growth rate vector; cap: carrying capacity vector; inter: N*N matrix of interaction weights between nodes; dt: Float or vector of sizes of time steps when simulating the continuous-time dynamics; stochastic: Boolean determining whether to simulate the stochastic or deterministic dynamics; pertb: Vector of perturbation magnitude of nodes' growth`) | Knebel et al., 2013 |
|---|---|---|
| Sherrington-Kirkpatrick dynamics (`simulate_sherrington`) | Simulates Sherrington-Kirkpatrick process on an Ising model, where each network node has two possible states, and switches between them based on long- and short-range interactions determined by the edges (**APs:**`- noise: True or false value to generate noise`) | Dorogovtsev et al., 2008 |
| Single random walk (`single_unbiased_random_walker`) | Simulates the dynamics of a single random walker on a network (**APs:**`- initial_node starting node for walk`) | Dorogovtsev et al., 2008 |
| Voter model dynamics (`voter`) | Creates voter model style dynamics where nodes are assigned one of two possible states at random and they update their state based on the state of its neighbors, with possibility of noise introduction | Gleeson et al., 2013 |

| | (**APs:**- noise: if noise is present, with this probability a node's state will be randomly redrawn from (-1,1) independent of its neighbors' states. If 'automatic', set noise to 1/N.) | |

**Similar packages and works**

Here we bring a comprehensive network reconstruction, comparison and dynamics package to the R environment. Some similar packages already exist in other languages. We would especially like to mention *netrd* package in Python (McCabe et al., 2020), which inspired and structured our current framework and functionality in R, for giving the same suite of tools to the ecological community, which is dominantly R-using.

Other works have aggregated methods from diverse sources for information-theoretic methods for time series network inference in Java (Lizier, 2014), and linear and non-linear causal discovery network algorithms for time-series data in Python (Runge et al., 2019). For network comparison, the R package *NetworkDistance* has 12 graph distance metrics (see You, 2019). We provide methods which have not necessarily been covered in this package, to increase the breadth of methods available in the R environment.

Providing this unified R package suite for inferring, comparing and simulating dynamics in networks will help the ecological community, obviating the need to scour multiple locations for different methods and their implementations.

**Software and code availability**

Three software packages described in this work are available at https://github.com/Fagan-Lab/

**Acknowledgements**

We would like to thank Morelle Tchuindjo and Nathan Stiff for their contributions to the project. We acknowledge the role of *netrd* package in Python in providing a baseline for our R development at all times. AS thanks National Science Foundation award DGE-1632976 for training and support.

**Author contributions**

AS and WFF conceptualized the project and wrote the manuscript; AS, TB, ZZ, LP, and ML contributed to the package development process; All authors provided feedback and approved the publication.